\documentclass[final]{ring20} 
\usepackage{amsmath}
\usepackage{amsfonts}
\DeclareMathOperator{\argmax}{arg\,max}
\DeclareMathOperator{\argmin}{arg\,min}

\newcommand{\xx}{\mathbf{x}}
\newcommand{\dd}{\mathbf{d}}
\def\permille{\ensuremath{{}^\text{o}\mkern-5mu/\mkern-3mu_\text{oo}}}

\graphicspath{%
{./Figures/}%
}

\title{Bayesian statistical analysis of hydrogeochemical data using point processes: a new tool for source detection in multicomponent fluid mixtures}

\author[1]{C. Reype}
\author[2]{A. Richard}
\author[1]{M. Deaconu}
\author[3]{R. S. Stoica}

\affil[1]{Universit\'e de Lorraine, CNRS, Inria, IECL, F-54000 Nancy, France}
\affil[2]{Universit\'e de Lorraine, CNRS, GeoRessources, F-54000 Nancy, France}
\affil[3]{Universit\'e de Lorraine, CNRS, IECL, F-54000 Nancy, France}

\shortauthor{C. Reype et al.} %
\shorttitle{Bayesian analysis of geological data using point processes}

\begin{document}
\maketitle

\begin{abstract}
Hydrogeochemical data may be seen as a point cloud in a multi-dimensional space. Each dimension of this space represents a hydrogeochemical parameter (\textit{i.e.} salinity, solute concentration, concentration ratio, isotopic composition...). While the composition of many geological fluids is controlled by mixing between multiple sources, a key question related to hydrogeochemical data set is the detection of the sources. By looking at the hydrogeochemical data as spatial data, this paper presents a new solution to the source detection problem that is based on point processes. Results are shown on simulated and real data from geothermal fluids.
\end{abstract}

\section*{Introduction}
The composition of many geological fluids is controlled by variable contributions of multiple sources (\textit{e.g.} seawater, meteoric water, hydrothermal water). The knowledge of these sources helps to built conceptual and quantitative models of fluid and mass transfer in the Earth's crust \citep{YarBod14}. If the sources are known, the contribution of each sources in every mixture can be inferred from hydrogeochemical data \citep[\textit{e.g.}][]{CarVaz04,SkuLon15}. In the case where the sources are not known, they can be inferred from the data \citep[\textit{e.g.}][]{Pint20}. 

The paper presents a Bayesian method of source detection based on point processes. The method is inspired by pattern detection methodologies used in image analysis, animal epidemiology and astronomy~\citep{StoDesZer03,StoiGayKret07,StoMaSa05,StoMaSa07}.

\section{Materials and methods}
\label{Method}

Let $\xx = \{x_i, i=1,\dotsc,n\}$ be a set of sources, giving the source position within a multi-dimensional (in practice two-dimensional) space formed by the hydrogeochemical parameters. A data point $d$ is a mixture of these sources (\textit{i.e.} it is explained by these sources) if it is a barycenter of these sources as stated in \citep{Fau97} \textit{i.e.}
\begin{equation}
    d=\sum_{i=1}^n \gamma_i x_i
\end{equation}
with $0\leq \gamma_i\leq 1$ for each $i$ and $\sum_{i=1}^n \gamma_i=1$.

In the Euclidean plane, the source pattern (\textit{i.e.} set of sources) is unknown and also somehow outlined by the set of hydrogeological data points. The key hypothesis at the basis of our work is that this pattern is made of interacting points. A preliminary condition for our model is that the hidden sources pattern exhibits the following properties~:
\begin{itemize}
\item the number of sources is not known but it should be controlled or minimal in a certain sense
\item two sources cannot be too close
\item the data points originating from a mixture of sources should be rather close to them
\item the convex hull enclosing the set of data points is enclosed within the convex hull given by the source positions.
\end{itemize}

These hypotheses allow to consider the sources as a realisation of a point process described by a Gibbs probability density:
\begin{equation*}
p(\xx|\theta) = \frac{\exp\left[-U(\xx|\theta)\right]}{Z(\theta)} = \frac{\exp\left[-U_{\dd}(\xx|\theta) - U_{i}(\xx|\theta)\right]}{Z(\theta)}
\end{equation*}
with $\xx$ the configuration of sources (or set of sources), $Z(\theta)$ the normalising constant and $U$ the energy function. 

The energy function is built as the sum of two components. The first term, $U_{\dd}(\xx|\theta)$ is the data term and it controls the positioning of the sources with respect to the observed data points $\dd=\{d_1,d_2,\ldots,d_m\}$. Its expression is given by 
\begin{equation*}
U_{\dd}(\xx,\theta) = \theta_1 g(\xx,\dd) + \theta_2 \sum_{j=1}^m \alpha(d_j,\xx) + \theta_3 n_e(\xx,\dd) .
\end{equation*}
Here $g(\xx,\dd)$ is the absolute value difference between the area of the sources and the data point convex hull, respectively. The function $\alpha(d_j,\xx)$ represents the minimum distance between the data point $d_j$ and the sources cloud and it is given by $\min\{\|d_j-x_i\|_2^2:\,i=1,\dots,n(\xx)\}$, with $n(\xx)$ the number of sources in the configurations. The measure $n_e(\xx,\dd)$ counts the number of data points in $\dd$, that belong to the convex hull given by $\xx$. The parameters $\theta_1, \theta_2 \geq 0$ and $\theta_3 \leq 0$ are chosen such that to penalize important differences between the convex hull areas, to encourage the source to be situated rather close to the data points and to increase the number of data points that are explained by the sources, respectively.

The second term, $U_{i}(\xx|\theta)$ is the interaction term and it writes as 
\begin{equation*}
U_{i}(\xx,\theta) = \theta_4 n(\xx)+\theta_5 n_{r}(\xx) .
\end{equation*}
with  $n_r(\xx)$ the number of pairs of sources at distance shorter than $r$, which is a pre-fixed known value. The parameters $\theta_4, \theta_5 \geq 0$ are chosen in order to penalize a too high number of sources and pairs of sources situated too close, respectively.

The point process on a finite domain $W$ (\textit{i.e.} $\mu(W)=\int_W \xi\,d\xi < \infty$), that is defined by the previous energy function, is well defined and locally stable~\cite{Sto14}. 
Based on these terms, the model is able to generate point configurations that exhibit the properties required by the assumed hypotheses. The source pattern is estimated by the point configuration that maximises the probability density $p(\xx|\theta)$
\begin{equation}
\widehat{\xx} = \argmax_{\xx \in \Omega} p(\xx|\theta) = \argmin_{\xx \in \Omega}U(\xx|\theta).
\label{eq:sourcesSolutions}
\end{equation}
The solution of the problem~\eqref{eq:sourcesSolutions} is obtained by implementing a simulated annealing algorithm. This algorithm is a global optimisation method that iteratively samples from $p(\xx|\theta)^{1/T}$ while making $T \rightarrow 0$ slowly. Convergence properties of this algorithm are shown in~\cite{StoiGregMate05}.

\subsection{Optimisation algorithm}
The implemented simulated annealing algorithm has the following structure:

\begin{itemize}
\item[1)] set $\theta$, $T$, $k_{max}$, $\xx^{(0)}$, $c$ and $k=1$
\item[2)] while $k\leq k_{max}$
\begin{itemize}
\item[a)]generate $\xx^{(k)}$ with probability $p(\xx^{(k-1)}|\theta)^{1/T}$
\item[b)]set $T=c*T$ and $k=k+1$
\end{itemize}
\item[3)] set $\xx=\xx^{(k)}$
\end{itemize}

This structure implements a sub-optimal cooling schedule, for practical reasons. An optimal logarithmic cooling schedule as specified by~\cite{StoiGregMate05} may be considered.

The sampling of $p(\xx|\theta)$ is done via the Metropolis-Hasting algorithm described below~:

\begin{itemize}
\item[1)] set $r_c$, $p_b$, $p_d$, $p_c$ with $p_b+p_d+p_c\leq1$
\item[2)] with probability $p_b$ choose birth, with probability $p_d$ choose death and with probability $p_c$ choose change. 
\begin{itemize}
\item[birth:]\begin{itemize}
\item[a)] generate a random point $\eta$ on $W$ and set $\xx'=\xx \cup \{\eta\}$
\item[b)] calculate $\beta_{b}=\min\{1,\frac{p_d}{p_b}\frac{p(\xx \cup \{\eta\}|\theta)}{p(\xx|\theta)}\frac{\mu(W)}{n(\xx)+1}\}$
\end{itemize}
\item[death:]\begin{itemize}
\item[a)] choose a point $\eta$ of $\xx$ and set $\xx'= \xx \setminus \{\eta \}$
\item[b)] calculate $\beta_{d}=\min\{1,\frac{p_b}{p_d}\frac{p(\xx \setminus \{\eta\}|\theta)}{p(\xx|\theta)}\frac{n(\xx)}{\mu(W)}\}$
\end{itemize}
\item[change:]\begin{itemize}
\item[a)] choose a point $\eta$ of $\xx$ and generate a random point $\xi$ in  the ball $B(\eta,r_c)$ and set $\xx'= \xx \setminus \{\eta\}\cup\{\xi\}$
\item[b)] calculate $\beta_{c}=\min\{1,\frac{p(\xx\setminus \{\eta\}\cup\{\xi\}|\theta)}{p(\xx|\theta)}\}$
\end{itemize}
\end{itemize}
\item[3)] the new configuration $\xx=\xx'$ is accepted with the appropriate probability $\beta$ ; otherwise the algorithm remains in the same state $\xx$.
\end{itemize}

The previous dynamic is $\phi-$ irreducible, Harris recurrent and geometric ergodic, guaranteeing the convergence of the algorithm towards the distribution of interest given by $p(\xx|\theta)$~\cite{Lies00,MollWaag04,Sto14}.

\section{Results}
\label{sec:appli}
The proposed model was tested on two different data sets. The first one is a simulated data set, the second one is a set of hydrogeochemical data from geothermal fluids described in \citep{Pint20}. The model is coded in C++, and the results are displayed in R with the library "ggplot".

We set $T=1000$, $k_{max}=10000$, $c=0.995$, $r_c=0.3$, $p_b=0.35$, $p_d=0.35$ and $p_c=0.3$. The parameters $\theta$ were chosen separately, for each data set, after several trials and errors.

\subsection{Simulated data}
The data are created by generating three sources. The vector of contributions of each sources to a data point is generated by a Dirichlet law with parameters $(1,1,1)$. Hence the data points are points uniformly distributed in the convex hull given by the sources positions in the 2D space of hydrogeochemical parameters. Moreover, a Gaussian noise, of mean $0$ and variance $10^{-1}$ for each coordinate, was added to each datapoint to represent the noise during the measurement.

\begin{figure}[htbp]
\centering\includegraphics[scale=0.4]{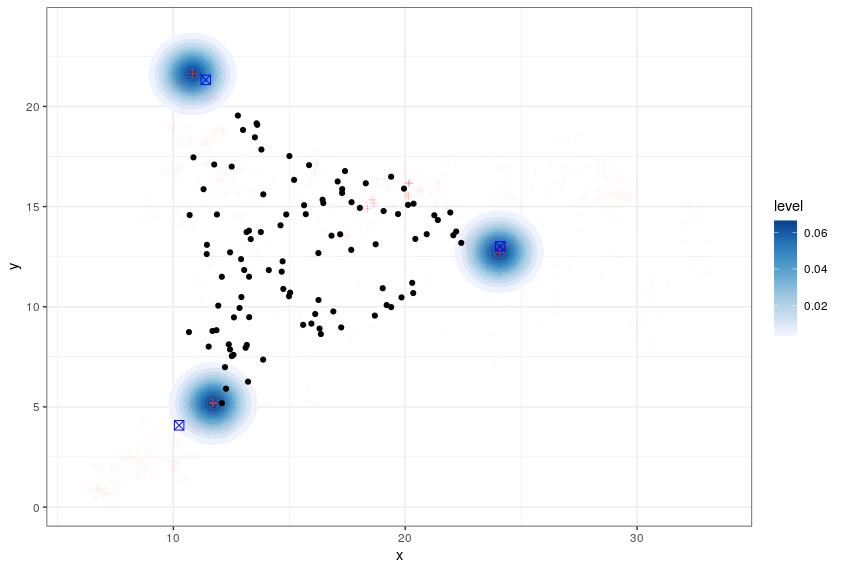}
\caption{Point process source detection for the simulated data in the case of a three-component fluid mixing system where $x$ and $y$ are the concentrations of two solutes (arbitrary units)}
\label{fig: sim}
\end{figure}

We set the parameters to $\theta=(100,1,-100,700,50)$ and $r=2$. The results are shown in Figure \ref{fig: sim}. The black dots are the data points, the blue symbols are the real sources and the gradient of blue color shows the density of simulated sources.

There are three areas that exhibit a high density of simulated sources, which corresponds to the actual number of sources. Moreover their positions are really close to the real sources.

\subsection{Real data}
\begin{figure}[htbp]
\centering\includegraphics[scale=0.4]{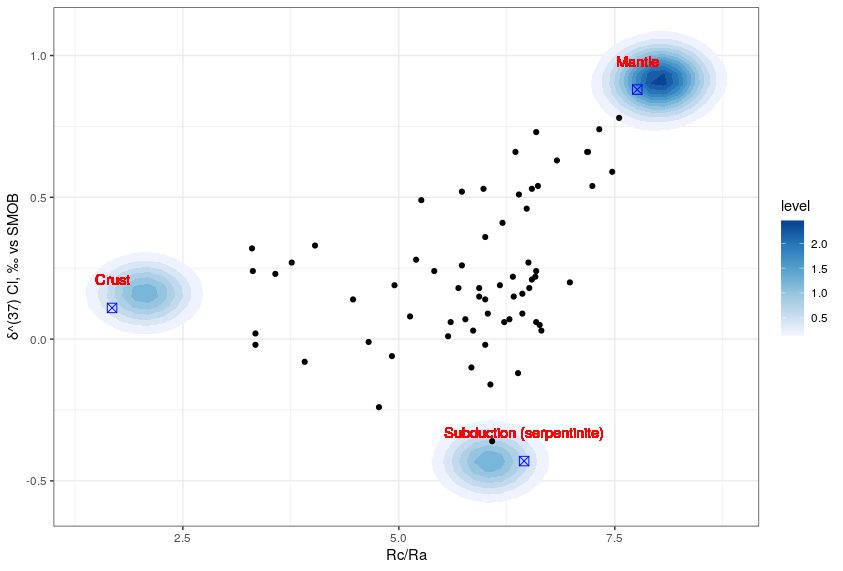}
\caption{Point process source detection for a three-component geothermal fluid mixing system where $\delta^{37}$Cl, \permille vs SMOC and Rc/Ra are respectively the stable isotopic composition of chlorine and the $^4He/ ^3He$ ratio of the samples (data from \citep[Figure 4]{Pint20})}
\label{fig: fig4}
\end{figure}

\begin{figure}[htbp]
\centering\includegraphics[scale=0.4]{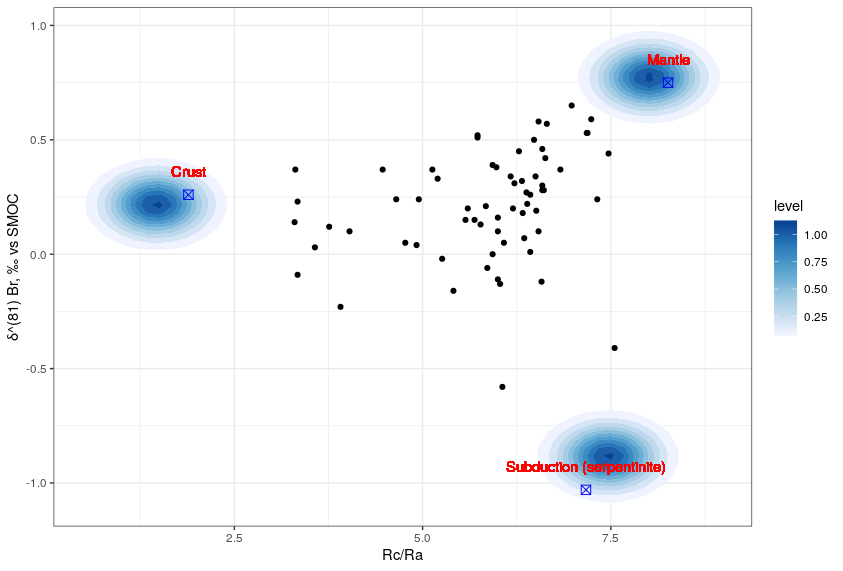}
\caption{Point process source detection for a three-component geothermal fluid mixing system where $\delta^{81}$Br, \permille vs SMOB and Rc/Ra are respectively the stable isotopic composition of bromine and the $^4He/ ^3He$ ratio of the samples (data from \citep[Figure 5]{Pint20})}
\label{fig: fig5}
\end{figure}

We are now comparing the results of our model with the results of the model presented in \citep{Pint20}. This model gives the smallest triangle (in term of area) that enclose the data. In Figure \ref{fig: fig4} we set $\theta=(200,1,-1200,210,5)$, and in Figure \ref{fig: fig5} we set $\theta=(196,2,-1200,220,5)$. The parameters are completely different than in \ref{fig: sim} because the data are not in the same scale.

The model is able to detect the number and the position of the sources inferred in ~\citep{Pint20}, while the model sufficient statistics provides a more complete morpho-statistical description of the sources. 

\section{Conclusions and perspectives}
\label{sec:discussion}
Clearly, the use of this method requires at least partial knowledge regarding the model parameters. Such a knowledge is built by embedding the available geological information into prior distributions. 

This new tool should be improved in order to become helpful not only in the analysis of geological fluids but also in other fields that deals with mixtures \citep[\textit{e.g.}][]{PhiGre03,LonVerErs18}. This is possible due to the use of the embedded spatial and Bayesian paradigms.

\section*{Acknowledgments}

This work was performed in the frame of the DEEPSURF project ( http://lue.univ-lorraine.fr/fr/impact-deepsurf ) at Universit\'e de Lorraine. This work was supported partly by the french PIA project Lorraine Universit\'e d'Excellence, reference ANR-15-IDEX-04-LUE.


\bibliography{bibliographie}

\end{document}